\documentclass[showpacs,amsmath,amssymb,onecolumn]{revtex4}
\usepackage{pstricks,amsmath,bbm,mathrsfs,amssymb,psfrag,pifont,times,mathptmx}
\usepackage[utf8]{inputenc}
\usepackage{subfigure}
\usepackage[english]{babel}
\usepackage{color}
\usepackage{array}
\usepackage[draft]{graphicx}

\newcolumntype{P}[1]{>{\centering\arraybackslash}p{#1}}

\begin{document}
\title{Optimizing Silicon photomultipliers for Quantum Optics}
\author{Giovanni Chesi}
\affiliation{Department of Science and High Technology, University of Insubria, Via Valleggio 11, I-22100 Como (Italy)}
\author{Luca Malinverno}
\affiliation{Department of Science and High Technology, University of Insubria, Via Valleggio 11, I-22100 Como (Italy)}
\author{Alessia Allevi}\email{alessia.allevi@uninsubria.it}
\affiliation{Department of Science and High Technology, University of Insubria, Via Valleggio 11, I-22100 Como (Italy)}
%\affiliation{Institute for Photonics and Nanotechnologies, CNR, Via Valleggio 11, I-22100 Como (Italy)}
\author{Romualdo Santoro}
\affiliation{Department of Science and High Technology, University of Insubria, Via Valleggio 11, I-22100 Como (Italy)}
\author{Massimo Caccia}
\affiliation{Department of Science and High Technology, University of Insubria, Via Valleggio 11, I-22100 Como (Italy)}
\author{Alexander Martemiyanov}
\affiliation{Department of Science and High Technology, University of Insubria, Via Valleggio 11, I-22100 Como (Italy) \\ ITEP, Bolshaya Cheremushkinskaya 25, 117218, Moscow (Russia)}
\author{Maria Bondani}
\affiliation{Institute for Photonics and Nanotechnologies, CNR, Via Valleggio 11, I-22100 Como (Italy)}
\date{\today}
%%%%%%%%%%%%%%
\begin{abstract}
Silicon Photomultipliers are potentially ideal detectors for Quantum Optics and Quantum Information studies based on mesoscopic states of light. However, their non-idealities hampered their use so far. An optimal mode of operation has been developed and it is presented here, proving that this class of sensors can actually be exploited for the characterization of both classical and quantum properties of light.
\end{abstract}
\maketitle

%%%
\section{INTRODUCTION}\label{sec:intro}
Nowadays, Quantum Technologies are receiving a boost due to their potential impact in the field of information processing. In particular, photonics represents an ideal platform thanks to the high transmission rates and the robustness of optical states against decoherence effects. In this context, the development of new light sources and detectors, more compact and versatile, is a priority. For what concerns detection, the last two decades have seen the implementation of novel classes of photodetectors endowed with photon-number-resolving capability.
Among them, Silicon photomultipliers (SiPMs) \cite{akindinov,bondarenko,saveliev,piemonte,renker} have received only a moderate attention in the field of Quantum Optics since their non-idealities have till now prevented their use in revealing quantum light features \cite{afek,kala12}. On the other hand, the use of SiPMs would be highly desirable due to their compactness, robustness and cost.\\
SiPMs are arrays of cells connected in parallel to a common output. Each cell is a p-n junction operated above the  breakdown voltage so that a single carrier is likely to trigger a Geiger-M$\rm \ddot{u}$ller avalanche multiplication by impact ionization. SiPMs feature single-photon sensitivity and an unprecedented photon-number-resolving capability, providing information about the intensity of the incoming light by counting the number of fired cells. The quantum efficiency of the new SiPMs can reach a maximum of 60$\%$ in the blue spectral region. \\ 
SiPMs are affected by spurious stochastic effects: dark count, cross talk and after pulses. 
Frequency of dark counts, namely spurious avalanches triggered by  thermally generated charge carriers, increases with bias voltage and temperature.
Optical cross talk \cite{gola} arises since electrons accelerated in the avalanche process produce secondary photons that may trigger avalanches in a neighboring cell. While this phenomenon was significant (of the order of 20$\%$) in the first generations of SiPMs, it has been decreased to a few percent in the latest generation, in which metal-filled trenches around the cells inhibit photon propagation.
SiPMs can be also affected by after pulses, which are defined as triggered avalanches due to the release of trapped carriers captured in previous avalanches. In the new generation of these detectors after-pulsing is negligible.\\ 
%Moreover, since this effect occurs at times longer than the duration of the light signal, it does not affect the detector response when the information is gathered by the peak of the sensor signal.\\ 
In this paper, we consider the latest generation of SiPMs and investigate to which extent the non-idealities could be detrimental for the effective exploitation of such detectors in Quantum Optics. 
By testing classical states of light we show that a comprehensive theoretical model, in which dark counts and cross talk are included, is in excellent agreement with the values of statistical moments and distributions obtained from the experimental data. We demonstrate that by a proper pulse shaping, digitization and processing, the impact of the spurious effects can be made negligible. 
In particular, by means of a peak-and-hold acquisition system we are able to properly reconstruct the photon-number statistics of the measured states without the need of taking into account the non-idealities. These results are crucial for the observation of non-classical features of light. 

\section{THE DETECTOR MODEL} \label{sec:model}
As mentioned in Section~\ref{sec:intro}, SiPMs are characterized by the presence of dark-count and cross-talk effects. The occurrence of dark counts is a stochastic process that in the temporal domain is randomly distributed according to a Poissonian distribution. This means that when the detector output is integrated over a gate synchronous with the light source, the mean value of dark counts depends linearly on the gate width.\\
The dark-count rate (DCR) can be directly obtained by measuring, in the absence of light, the frequency of the detection events that are above the electronic noise threshold.
%The result of a threshold scan of dark counts is displayed in Fig.~\ref{staircase}.}
\begin{figure}[htbp]
\begin{center}
\includegraphics[width=0.8\columnwidth]{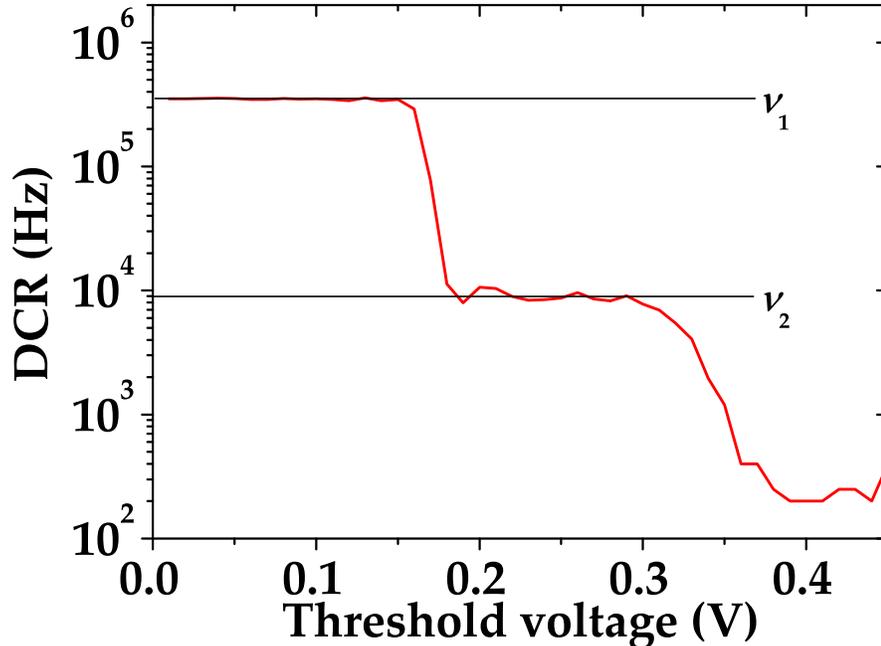}
\vspace{-1cm}
\end{center}
\caption{(Color online) Threshold scan of SiPM performed at room temperature with no impinging light.}
\label{staircase}
\end{figure}
Optical cross-talk events \cite{piemonte,du,buzhan} can occur simultaneously to the light signal (prompt cross talk) or at a later time (delayed cross talk). The first ones occur when the Geiger M$\ddot{\rm u}$ller-avalanche-generated photons are directly absorbed into the active layer of the cell, thus triggering an avalanche simultaneous with the primary absorbed photons. There is no possibility to discriminate between primary detection events and cross-talk events, since the last ones contribute to the signal as additional fired cells. Nevertheless, the rate of prompt cross-talk events can be directly estimated as the ratio
\begin{equation}
\epsilon = \nu_2/\nu_1,
\end{equation}
where $\nu_1$ is the DCR measured by setting the threshold at half the amplitude of the signal corresponding to a single avalanche, and $\nu_2$ is the DCR when the threshold is set to one and half single-cell signal \cite{RamilliXT}. In order to obtain the values of $\nu_1$ and $\nu_2$, DCRs have been measured by scanning the threshold values (Fig. \ref{staircase}).
As to delayed cross-talk events, if a photon generated by the GM avalanche is absorbed in the epitaxial layer, parasitic electric field could move it into the active layer and trigger an avalanche at a later stage. The temporal development of delayed cross-talk probability has been investigated in Ref.~\cite{nagy14}, where an empirical model for the cross-talk probability density function, $\delta \epsilon (t)$, has been proposed:
\begin{equation}\label{XCtemp}
\delta \epsilon(t) = a \exp\left(- \frac{t}{\tau_{\rm xc}} \right).
\end{equation} 
In Eq.~(\ref{XCtemp}), $a$ is a constant, and $\tau_{\rm xc}$ is a characteristic time. 
%The expression in Eq.~(\ref{XCtemp}) can be interpreted as a cross-talk probability per unit time. 
This model correctly accounts for the behavior of cross talk at times shorter than the temporal development of the detector signal.  
The integral of Eq.~(\ref{XCtemp}) over different choices of gate widths yields the effective cross-talk probability $\epsilon$ to be used in the model for detection. In the following Section, we will show that this model is consistent with our data.\\
In order to interpret the experimental data, we have developed a comprehensive model for SiPMs \cite{ramilli}.
First of all, we assume that the detection process of the entire SiPM array is described by a Bernoullian distribution, $B_{m,n}(\eta)$, so that the distribution of photoelectrons, ${\rm P_{el}}(m)$, can be written as a function of the distribution of photons, ${\rm P_{ph}}(n)$, through
\begin{equation}
{\rm P_{el}}(m) = \sum_{n=m}^\infty B_{m,n}(\eta) {\rm P_{ph}}(n) = \sum_{n=m}^\infty {{n}\choose{m}} \eta^m (1- \eta)^{n-m} {\rm P_{ph}}(n),
\end{equation}
where $\eta$ is the detection efficiency, $n$ is the number of incident photons, and $m$ that of photoelectrons. 
Dark counts can be modelled by a Poissonian distribution
\begin{equation}\label{Pdc}
{\rm P_{dc}}(m) = \frac{(\langle m \rangle_{\rm dc})^m}{m!}\exp{(-\langle m \rangle_{\rm dc})},
\end{equation}
where $\langle m \rangle_{\rm dc}$ is the mean value of dark counts. 
Moreover, the optical cross-talk effect, which is a genuine cascade phenomenon, at first order can be written as \cite{afek}
\begin{equation}
C_{k,l}(\epsilon) = {{l}\choose{k-l}} \epsilon^{k-l}(1- \epsilon)^{2l-k},
\end{equation}
being $\epsilon$ the probability that the avalanche from a cell triggers one neighbor cell, $l$ is the number of photo-triggered avalanches and of dark counts, and $k$ is the resulting number of avalanches including cross talk.\\
Moreover, we assume that the amplification of the detector is described by a multiplicative parameter, $\gamma$ \cite{JMO}.
Thus, the photon-number distribution of the detector output is given by
\begin{equation}
{\rm P}(x_{\rm out}) = \gamma \sum_{m=0}^k C_{k,m}(\epsilon) \sum_{j=0}^m {\rm P_{dc}}(j) {\rm P_{el}}(m-j),
\end{equation}
where $x_{\rm out}$ is the single-shot output of the overall detection chain. According to this model, $\langle x_{\rm out} \rangle = \gamma \langle k \rangle = \gamma (1 + \epsilon) (\langle m \rangle + \langle m\rangle_{\rm dc})$ and $\sigma^2_{x_{\rm out}} = \gamma^2 \left[ (1+ \epsilon)^2 (\sigma^2_m + \langle m \rangle_{\rm dc}) + \epsilon (1-\epsilon) (\langle m \rangle + \langle m\rangle_{\rm dc})\right]$.

\section{EXPERIMENTAL RESULTS}

\subsection{The detection chain: processing and data acquisition}
\noindent The current investigation has been performed using two SiPMs (MPPC S13360-1350CS, \cite{Hama1}), produced by Hamamatsu Photonics \cite{Hama2}, each one having $667$ pixels in a $1.3 \times 1.3$ mm$^2$ photosensitive area, a pixel pitch equal to 50-$\mu$m, and a quantum efficiency of 40$\%$ at 460 nm.\\ 
%At the optimal operating voltage ($\sim 55$ V) and at room temperature ($\sim 25^{\circ}$ C), a dark-count rate $\sim 60$ kHz and a cross-talk $\epsilon \sim 3.5\%$ were measured, in good agreement with the datasheet of the detectors.\\
\begin{figure}[htbp]
\begin{center}
\includegraphics[width=0.8\columnwidth]{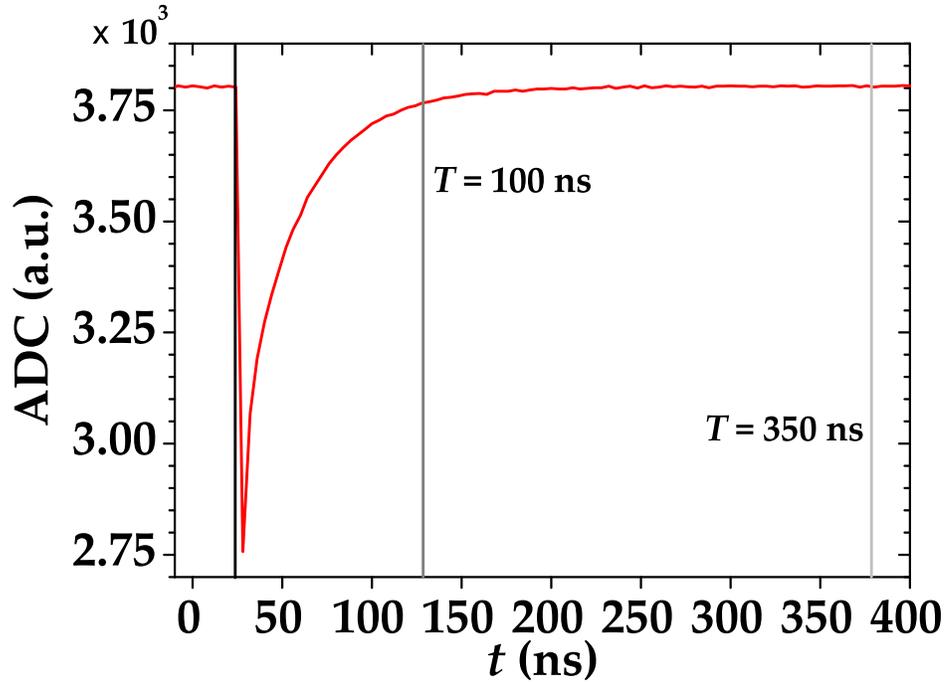}
\vspace{-1cm}
\end{center}
\caption{(Color online) A single-shot detector signal, acquired with the digitizer, is indicated as red curve together two possible integration gates, $T$. The gate ending with the vertical gray line is 100-ns long, whereas the one ending with the vertical light gray line is 350-ns long.}
\label{oscilloscope}
\end{figure}
Each detector output was sampled by a desktop waveform digitizer (DT5720, CAEN) operating at 12-bit resolution and at 250-MS/s sampling rate. The acquired traces were then integrated in post processing over different gate widths $T$, ranging from the whole waveform time evolution ($T \sim 350$ ns) down to the few-nanosecond-long peak duration. 
A typical digitized trace is shown in Fig.~\ref{oscilloscope} as red curve together with two possible integration gates, which are indicated by the vertical black line on one side and by the vertical gray or light gray line on the other side). 
\begin{figure}[htbp]
\begin{center}
\includegraphics[width=0.8\columnwidth]{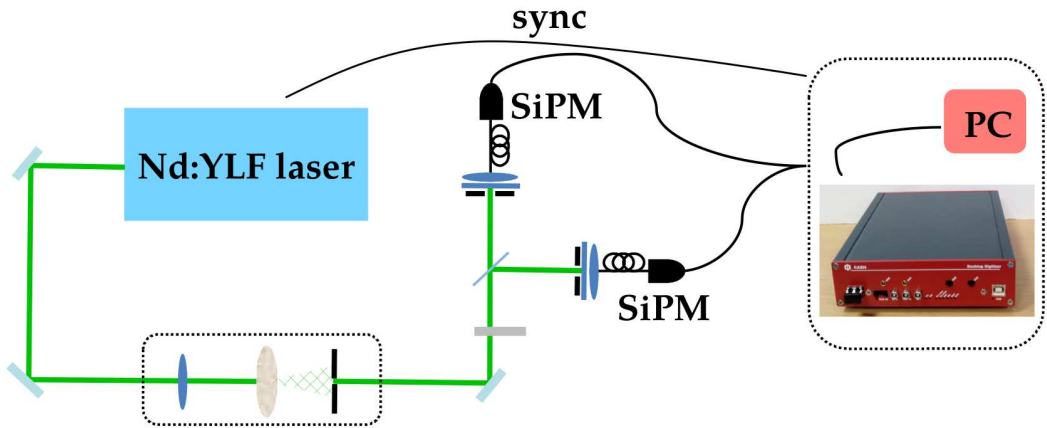}
\end{center}
\caption{(Color online) Experimental setup for the measurement of coherent and multi-mode thermal states. See the text for details.}
\label{setupclass}
\end{figure}
%First, an analog integration of the detector output was performed through two synchronous boxcar-gated integrators (SR250, Stanford Research Systems); second, 
%, so that the sampling is operated every 4 ns. 

\subsection{Experimental setup for the measurement of classical states of light}\label{classsetup}
\begin{figure}[htbp]
\begin{center}
\includegraphics[width=0.8\columnwidth]{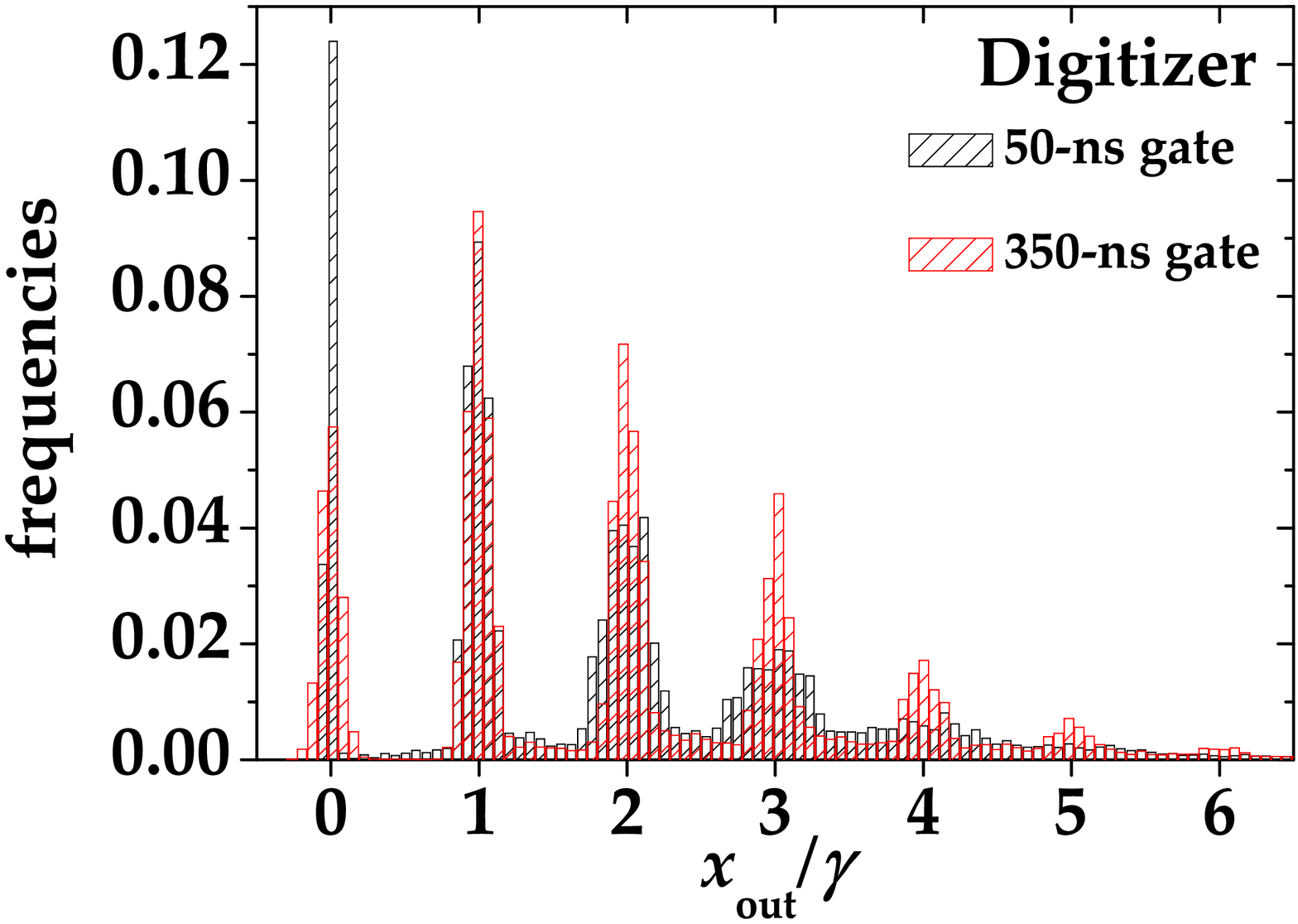}
\end{center}
\caption{(Color online) Normalized pulse-height spectra corresponding to a multi-mode thermal state with $\langle k \rangle \sim 2$ acquired with the digitizer and then integrated over different gate widths. Black: 50-ns gate; red: 350-ns gate.}
\label{PHSd}
\end{figure}

In order to investigate the performance of SiPMs, two light sources endowed with different photon-number distributions were investigated: Poissonian and multi-mode thermal statistics.
As shown in Fig.~\ref{setupclass}, the Poissonian light was provided by the second harmonic (at 523 nm) of a mode-locked Nd:YLF laser amplified at 500 Hz (High Q Laser), whereas the multi-mode thermal was obtained by passing the same laser beam through a rotating ground-glass disk and collecting nearly one single coherence area with an iris located $\sim$1 m apart from the disk \cite{arecchi}. Each light source was equally divided by means of a half-wave plate followed by a polarizing cube beam splitter, at whose outputs the two SiPMs were positioned. 
Multi-mode optical fibers with 600-$\mu$m-core-diameter were used to deliver the light to the sensors.
In order to collect the same light at different mean values, a variable neutral density filter wheel was used to attenuate the light from 0 to 2 optical densities. For each intensity, we recorded the response to $\sim$120,000 consecutive laser pulses. 
In Fig.~\ref{PHSd} we show two typical pulse height spectra corresponding to multi-mode thermal states with roughly the same mean number of photons ($\langle m \rangle \sim 2$). 
The signal was acquired with the digitizer and then integrated in post processing over 350-ns and 50-ns gate widths. We note that for the short gate the separation among peaks is worse than the one for the long gate, since a smaller number of samplings is considered. 
\begin{figure}[htbp]
\begin{center}
\includegraphics[width=0.8\columnwidth]{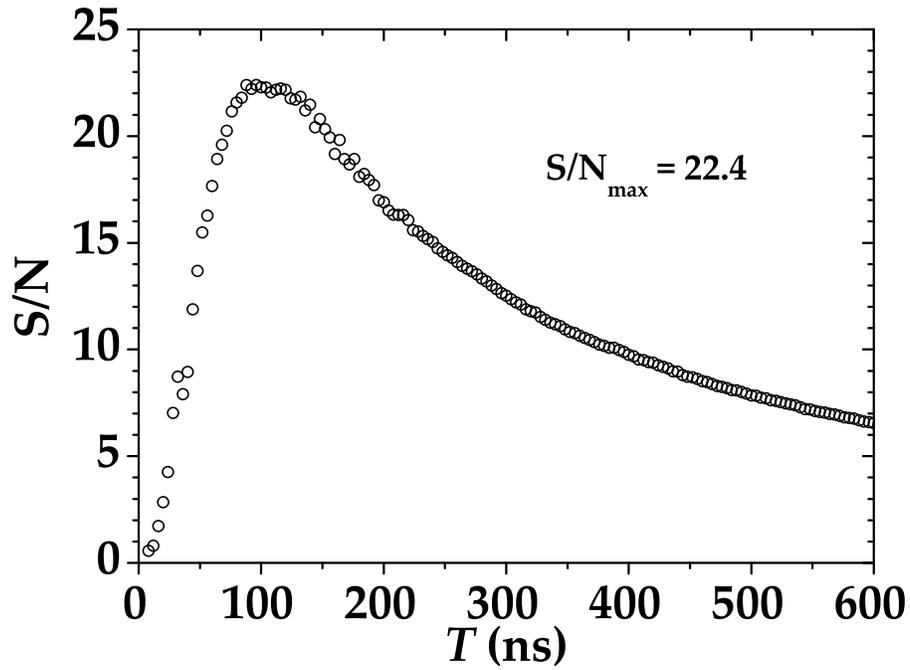}
\end{center}
\caption{(Color online) Signal-to-noise ratio S/N for the digitized signals in Fig.~\ref{PHSd} as a function of the integration time, $T$.}
\label{SN}
\end{figure}
In Fig.~\ref{SN} we plot the signal-to-noise ratio (S/N) corresponding to the light signal used to produce Fig.~\ref{PHSd} as a function of the integration time. The S/N is here defined as the ratio between the mean value of the integral of 1-photon peak and its variance. We note that the S/N increases at increasing values of the integration gate, up to a maximum value, which corresponds to the extinction of the signal, occurring at $T~\sim$ 150 ns (see Fig.~\ref{oscilloscope}). Note that similar values of S/N are achieved for very different settings of $T$ (see e.g. $T \sim 50$ ns and $T \sim 350$ ns), at which the measurement is differently affected by spurious stochastic effects. In the following we will show that the choice of the best integration width cannot be based on S/N values, at least for quantum optics applications.

%{figure}[htbp]
%\begin{center}
%\includegraphics[width=1\columnwidth]{phs_box}
%\end{center}
%\caption{(Color online) Normalized pulse-height spectra corresponding to a multi-mode thermal state with $\langle k \rangle \sim 2$ acquired with the boxcar and different integration gate widths. Black: 50-ns gate; magenta: 10-ns gate.}
%\label{PHSb}
%\end{figure}
%On the contrary, it is well evident from Fig.~\ref{PHSb}, in which some pulse-height spectra acquired with the boxcar integrators are shown, that in the case of an analog integration the better results are obtained with the smallest gate width, that is 10 ns. In fact, for this choice of gate the spurious effects due to delayed cross-talk events and dark counts are almost negligible, as shown hereafter.

\subsection{First- and second-order moments}
The analysis of the low-order moments of photon-number distributions, such as mean value and variance, can be sufficient to characterize the nature of light. For instance, it is well-known that the value of the Fano factor \cite{mandel}
\begin{equation}
F(n) = \frac{\sigma^2_n}{\langle n \rangle},
\end{equation}
in which $\sigma^2_n$ is the variance and $\langle n \rangle$ the mean number of photons,
is sufficient to discriminate among Poissonian, sub-Poissonian and super-Poissonian statistics, and hence can serve as an indicator of nonclassicality \cite{arimondo}.\\
In order to check the reliability of the SiPMs in detecting the characteristics of light states, we consider the first two moments of photon distributions, and use them to calculate the Fano factor. 
According to the model described in Sect.~\ref{sec:model} \cite{ramilli}, the Fano factor for the SiPM output is a linear function of the first moment $\langle x_{\rm out} \rangle$  
\begin{equation}
 F(x_{\rm out}) = \frac{\sigma^{(2)}_{x, \rm out}}{\langle x_{\rm out} \rangle} = \frac{Q_{\rm el+dc}} {\langle m \rangle + \langle m \rangle_{\rm dc}} \langle x_{\rm out} \rangle +\gamma \frac{1+3\epsilon}{1+\epsilon}, 
\label{Fano}
\end{equation}
where $Q_{\rm el+dc} = (\sigma^2_m + \langle m \rangle_{\rm dc}) / (\langle m \rangle + \langle m \rangle_{\rm dc}) -1$ is the Mandel factor of the detected light.\\

\noindent {\it Poissonian statistics}\\
The photon-number distribution of a coherent field is Poissonian, so that $\langle n \rangle = \sigma^2_n$. In this case, Eq.~(\ref{Fano}) reduces to
\begin{equation}                                                                                                                                                                      
F_{\rm coh}(x_{\rm out}) = \gamma \frac{1+3\epsilon}{1+\epsilon}.
\label{F} \\
\end{equation}
%Note that, in the case of Poissonian statistics, it is not possible to determine the amount of dark counts since they also feature Poissonian statistics (see Eq.~(\ref{Pdc})). 
%That is why in Eq.~(\ref{F}) only the amplification parameter and the cross-talk probability $\epsilon$ appear.
For the data we present, we decided to independently estimate the value of $\gamma$, as it corresponds to the distance between two consecutive peaks in the pulse-height spectrum and can be determined by means of a multi-peak fit of the spectrum histogram modelling each peak with a Gaussian function \cite{ramilli}.\\
%%%%%%%%%%%%%%%%%%%%%%%%%%%%%%%%%%%%%%
\begin{figure}[htbp]
\begin{center}
\hspace{-0.cm}
\includegraphics[width=0.8\columnwidth]{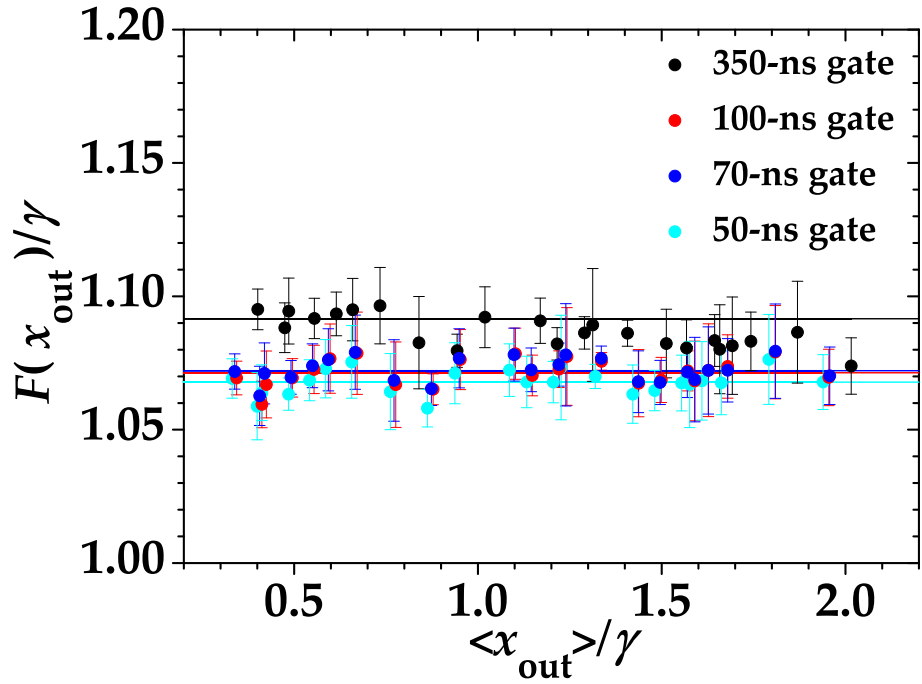}
%\hspace{-0.cm}
%\includegraphics[width=1\columnwidth]{grSkewnesscoher}
\end{center}
\caption{(Color online) Fano factor 
%(panel (a)) and Symmetry factor (panel (b)) 
for Poissonian statistics as a function of the mean output of the detector. Different colors correspond to different integration gate widths: black to 350-ns gate, red to 100-ns gate, blue to 70-ns gate, and cyan to 50-ns gate. 
Dots: experimental data; lines: fitting linear curves in which $\epsilon$ is the only fitting parameter. 
}
\label{FScoher}
\end{figure}
%%%%%%%%%%%%%%%%%%%%%%%%%%%%%%%%%%%%%%%%%
\noindent In Fig.~\ref{FScoher} we show the experimental values of the Fano factor as a function of the mean value of the output together with the linear fitting curves, in which $\epsilon$ is the only fitting parameter and $\gamma$ is fixed. All the data presented in the figure were obtained by the same dataset acquired with the digitizer and integrated in post processing over different gate widths. In detail, different colors correspond to different integration gate widths: black to 350-ns gate, red to 100-ns gate, blue to 70-ns gate, and cyan to 50-ns gate. In more detail, for the curves in Fig.~\ref{FScoher} we obtained the fit parameters shown in Tab.~\ref{tab1}.\\
\begin{table}[!htb]
\begin{center}
\renewcommand{\arraystretch}{1.5}
\begin{tabular}{|c|P{1.5cm}|P{2.5cm}|P{1.5cm}|}
\hline 
gate width & $\epsilon$ & CI($\epsilon$) & $\chi^2_{\nu}$ \\
\hline 
350 ns & 0.0480 & (0.0467, 0.0494) & 2.34 \\ 
\hline 
100 ns & 0.0370 & (0.0356,0.0383) & 0.68 \\ 
\hline 
70 ns & 0.0374 & (0.0359, 0.0390) & 0.60 \\ 
\hline 
50 ns & 0.0351 & (0.0336, 0.0366) & 0.75 \\ 
\hline 
\end{tabular} 
\caption{Values of the fitting parameter $\epsilon$ of the Fano factor as a function of the gate width in the case of coherent light. The symbol CI indicates the 95$\%$ confidence interval. In the last column the $\chi^2$ per degree of freedom is shown.}\label{tab1}
\end{center}
\end{table}
%For what concerns Fig.~\ref{FScoher}(b), we got $\epsilon = 0.0436 \pm 0.0005$ for 350-ns gate, $\epsilon = 0.0406 \pm 0.0004$ for 100-ns gate, $\epsilon = 0.0392 \pm 0.0004$ for 70-ns gate, and $\epsilon = 0.0389 \pm 0.0004$ for 50-ns gate. 
The horizontal linear behavior described by Eq.~(\ref{F}) is well supported by all the experimental data. By comparing the values of $\epsilon$ for each choice of gate width, we note that a larger integration gate results in a larger number of cross-talk-affected events, thus leading to a larger Fano factor. In particular, all the values of $\epsilon$ are comparable to that determined with the standard strategy described in Sect.~\ref{sec:model}, $i.e.$ $\epsilon =$ 0.03 $\pm$ 0.01. 
%On the contrary, the smallest investigated integration gate (i.e. $50$ ns) provides a value of $\epsilon$ more consistent with the data reported in the datasheet. 
The decrease of the cross-talk probability for small gate widths can be explained by considering the temporal distribution of cross-talk events, which, from Eq.~(\ref{XCtemp}), can be written as
\begin{equation} \label{epsvsgate}
\epsilon = \epsilon_0 + \int_0^T dt \delta\epsilon(t) = \epsilon_0 + \frac{\tau_{\rm xc}}{T} a \left[ 1-\exp \left( -\frac{T}{\tau_{\rm xc}} \right)\right],
\end{equation}
where $\epsilon_0$ is the prompt cross-talk probability and $T$ is the integration time (gate width).   
%%%%%%%%%%%%%%%%%%%%%%%%%%%%%%%%%%%%%%
\begin{figure}[htbp]
\begin{center}
\hspace{-0.cm}
\includegraphics[width=0.8\columnwidth]{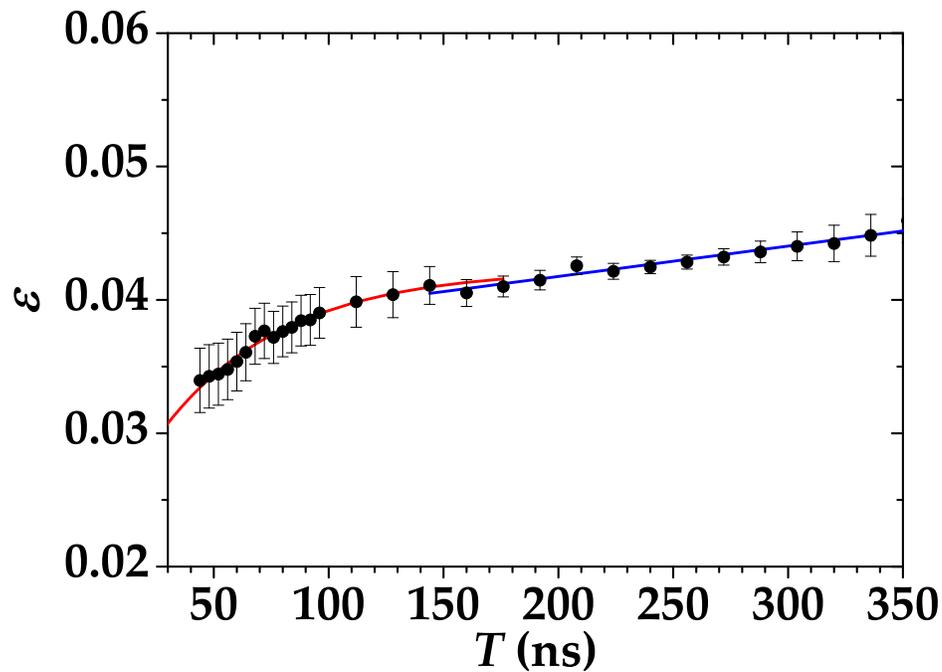}
\end{center}
\caption{(Color online) Cross-talk probability as a function of the length of the gate for $\langle m \rangle \sim 1.2$. Dots + error bars:experimental data; red curve: theoretical fitting curves for short and long times, respectively.
}
\label{grepsvsgate}
\end{figure}
%%%%%%%%%%%%%%%%%%%%%%%%%%%%%%%%%%%%%%%%%
In Fig.~\ref{grepsvsgate} we show the value of cross talk obtained from the calculated Fano factor  as a function of the gate width for a chosen mean value (dots + error bars). The experimental data exhibit two rather different trends for short and long times. For $T < 150$ ns, the data are well fitted by the model in Eq.~(\ref{epsvsgate}), while for longer gate widths the trend is linear. Noting that a 150-ns gate entirely covers the evolution of the signal, we can safely assess that the contribution at longer gates only comes from cross-tak events triggered by dark counts, which are linear in the gate width.
%. The expression in Eq.~(\ref{epsvsgate}) cannot account for both simultaneously, so we performed two separate fits.
By fitting the data in Fig.~\ref{grepsvsgate} up to $T = 150$ ns with Eq.~(\ref{epsvsgate}) we get: $\epsilon_0 = 0.0219 \pm 0.0024$, $a = 0.0004 \pm 0.0001$, and $\tau_{\rm xc} = 53 \pm 9$ ns. For longer times we fit the data with $\epsilon = m T + q$ and obtain $m= 2.2~10^{4} \pm 0.1~10^{4}$ Hz and $q = 0.0372 \pm 0.0004$. The term $m T$ can be interpreted as the cross-talk probability associated to dark-count events. 
%Note that these probability values are negligible as expected.
%Analogous conclusions for the value of $\epsilon$ as a function of the gate width can be drawn for the Symmetry factor. In particular, for each choice of gate width, the values of $\epsilon$ obtained by fitting the data corresponding either to $F$ and $S$ are very similar. 
\\
\\
\noindent {\it Multi-mode thermal statistics}\\
In the case of pseudo-thermal light impinging on the SiPM, the photon-number distribution is generally multi-mode thermal with $\mu$ modes equally populated. In this case, the Fano factor reduces to
\begin{equation}
F_{\rm mth}(x_{\rm out}) = \frac{1}{\mu} \left(1-\frac{\langle x_{\rm dc} \rangle}{\langle x_{\rm out} \rangle}\right)^2 \langle x_{\rm out} \rangle+\gamma \frac{1+3\epsilon}{1+\epsilon},
\label{Fmth}
\end{equation}
in which $\langle x_{\rm dc} \rangle = \gamma \langle m \rangle_{\rm dc}$ is the mean number of dark counts.
For multi-mode thermal light the Fano factor is a rational polynomial function, whose minimum value is $F_{\rm mth}(0) = \gamma (1+ \epsilon)/(1+ 3\epsilon)$, that is the Fano factor of coherent light.
%%%%%%%%%%%%%%%%%%%%%%%%%%%%%%%%%%%%%%%%%%%%%%%%%%
\begin{figure}
\begin{center}
\hspace{-0.cm}
\includegraphics[width=0.8\columnwidth]{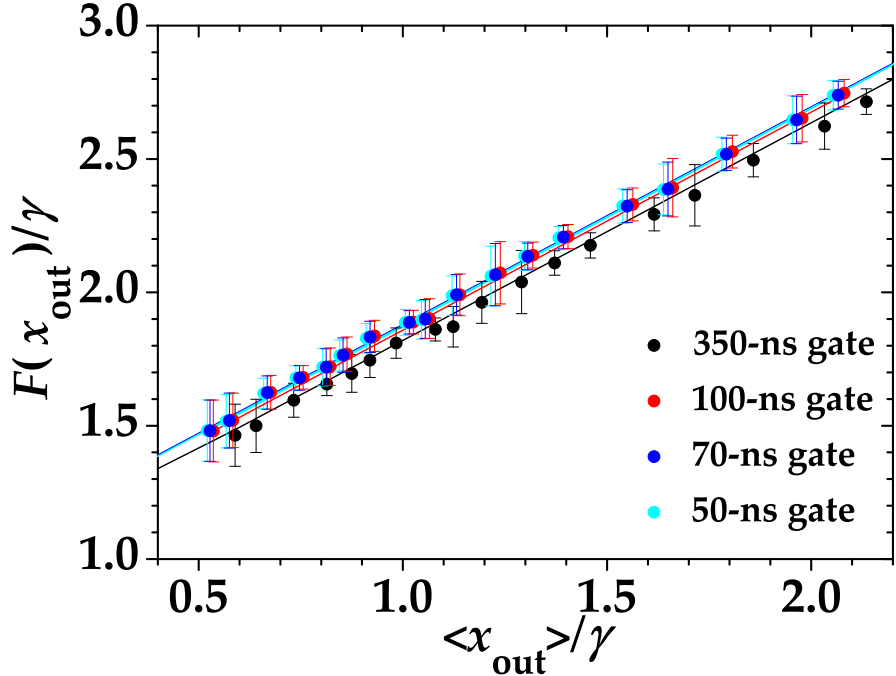}
%\hspace{-0.cm}
%\includegraphics[width=1\columnwidth]{grSkewnesstherm}
\end{center}
\caption{(Color online) Fano factor 
%(panel (a)) and Symmetry factor (panel (b))
 for multi-mode thermal statistics as a function of the mean output of the detector. Different colors correspond to different integration gate widths: black to 350-ns gate, red to 100-ns gate, blue to 70-ns gate and cyan to 50-ns gate. Dots: experimental data; lines: fitting curves with fitting parameters $\langle x \rangle_{\rm dc} $ and $\mu$, while $\epsilon$ is fixed from the plots in Fig.~\ref{FScoher}.
}
\label{FSmth}
\end{figure}
%%%%%%%%%%%%%%%%%%%%%%%%%%%%%%%%%%%%%%%%%%%%%%%%%
In Fig.~\ref{FSmth} the experimental values of Fano factor 
%(in panel (a)) and Symmetry factor (in panel (b)) 
are in good agreement with the fitting curves calculated according to Eq.~(\ref{Fmth}). 
%and (\ref{Smth})
We note that in the calculation of such curves, for each choice of the gate width (50 ns, 70 ns, 100 ns, and 350 ns), we used the values of cross-talk probability $\epsilon$ estimated in the case of coherent light. Also the factor $\gamma$ was separately estimated to further reduce the number of parameters. 
Since the different sets of data share the same parameters ($\mu$ and $\langle x \rangle_{\rm dc}$), for all the gate widths we performed a fitting procedure by imposing $\chi^2_{\nu} = 1$ and we assumed a scaling of $\langle x \rangle_{\rm dc}$ linear with the gate width. Moreover, since the light is the same for all the gate widths, we assumed that they shared the same value of $\mu$.
The obtained values are shown in Tab.~\ref{tab2}. We notice that the values of dark counts correspond to a DCR of $\sim 160$~kHz, which is quite similar to the rate independently measured with the standard technique presented in Sect.~\ref{sec:model}, $i. e.$ 140 kHz.
\begin{widetext}
\begin{center}
\begin{table}[!htb]
\renewcommand{\arraystretch}{1.5}
%\begin{footnotesize}
\begin{tabular}{|c|P{1.5cm}|P{2.8cm}|P{1.5cm}|P{1.5cm}|P{2.8cm}|}
\hline 
gate width & $\langle x_{\rm dc} \rangle$ & CI($\langle x_{\rm dc} \rangle$) & DCR (kHz) & $\mu$ & CI($\mu$) \\
\hline 
350 ns & 0.0563 & (0.0544, 0.0582) & 161 $\pm$ 11 & 1.2234 & (1.2225, 1.2243) \\ 
\hline 
100 ns & 0.0187 & (0.0168, 0.0206) & 187 $\pm$ 38 & 1.2234 & (1.2225, 1.2243) \\ 
\hline 
70 ns & 0.0086 & (0.0065, 0.0107) & 123 $\pm$ 60 & 1.2234 & (1.2225, 1.2243) \\ 
\hline 
50 ns & 0.0080 & (0.0078, 0.0082) & 160 $\pm$ 8 & 1.2234 & (1.2225, 1.2243) \\ 
\hline 
\end{tabular} 
\caption{Values of the fitting parameters $\langle x \rangle_{\rm dc}$ and $\mu$ of the Fano factor as functions of the gate width in the case of multi-mode thermal light. The symbol CI indicates the 95$\%$ confidence interval.}\label{tab2}
%\end{footnotesize}
\end{table}
\end{center}
\end{widetext}
%In addition, for Fig.~\ref{FSmth}(b) we achieved $\langle x_{dc} \rangle = 0.0518 \pm 0.0073$ and $\mu = 1.0859 \pm 0.0091$ for 350-ns gate, $\langle x_{dc} \rangle = 0.0000 \pm 0.0074$ and $\mu = 1.0803 \pm 0.0094$ for 70-ns gate, and $\langle x_{dc} \rangle = 0.0000 \pm 0.0073$ and $\mu = 1.0725 \pm 0.0092$ for 50-ns gate.
The values of the fitting parameters prove that, as expected, a larger integration gate width leads to a larger occurrence of non-idealities. Indeed, not only the cross-talk probability increases, but also the mean number of dark counts.
%%%%%%%%%%%%%%%%%%%%%%%%%%%%%%%%%%%%%%%%%%%%%%%%%%  
\subsection{Statistics}
\begin{figure}[htbp]
\begin{center}
\hspace{-0.25cm}
\includegraphics[width=0.5\columnwidth]{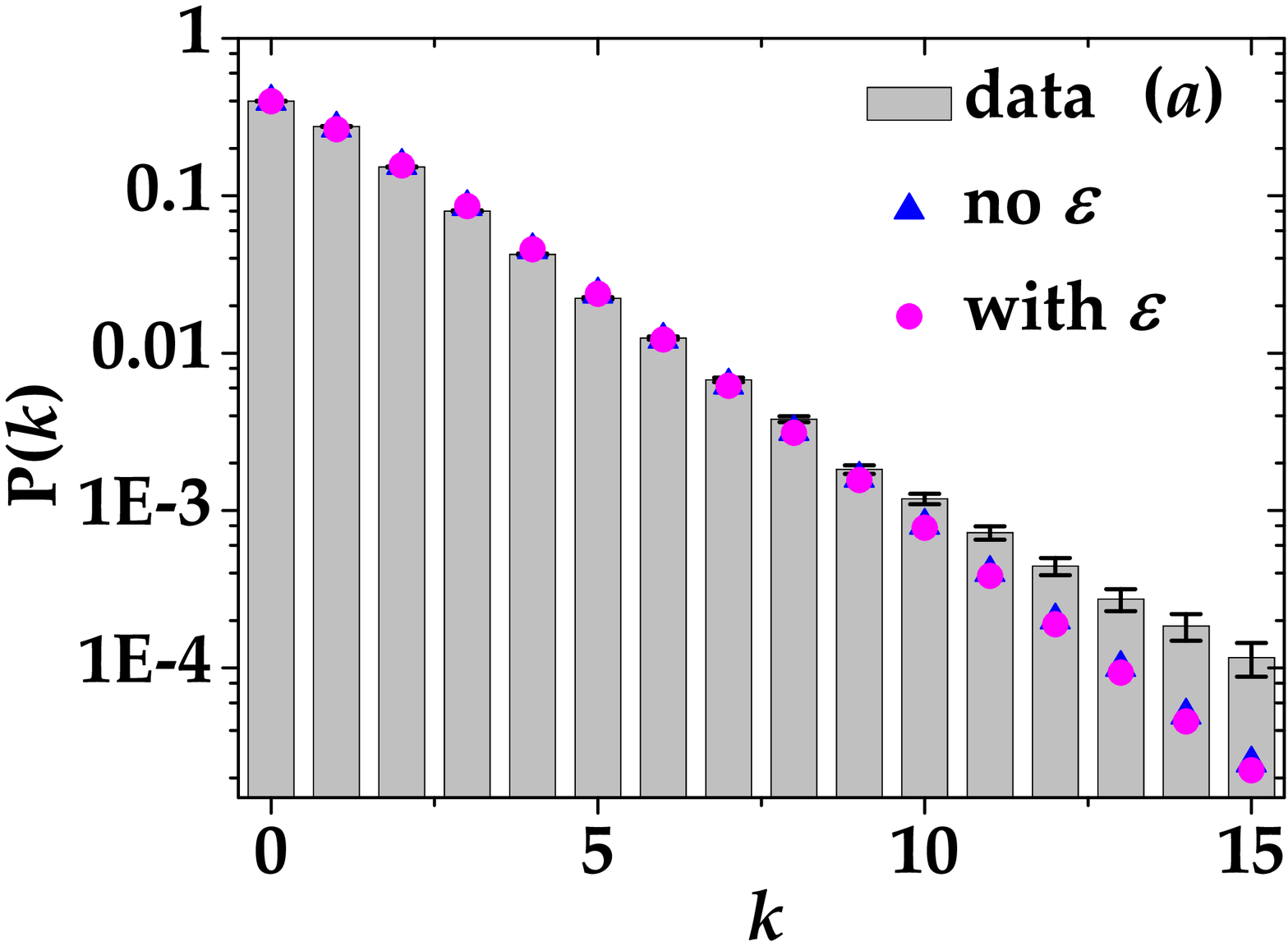}
\hspace{-1.6cm}
\includegraphics[width=0.5\columnwidth]{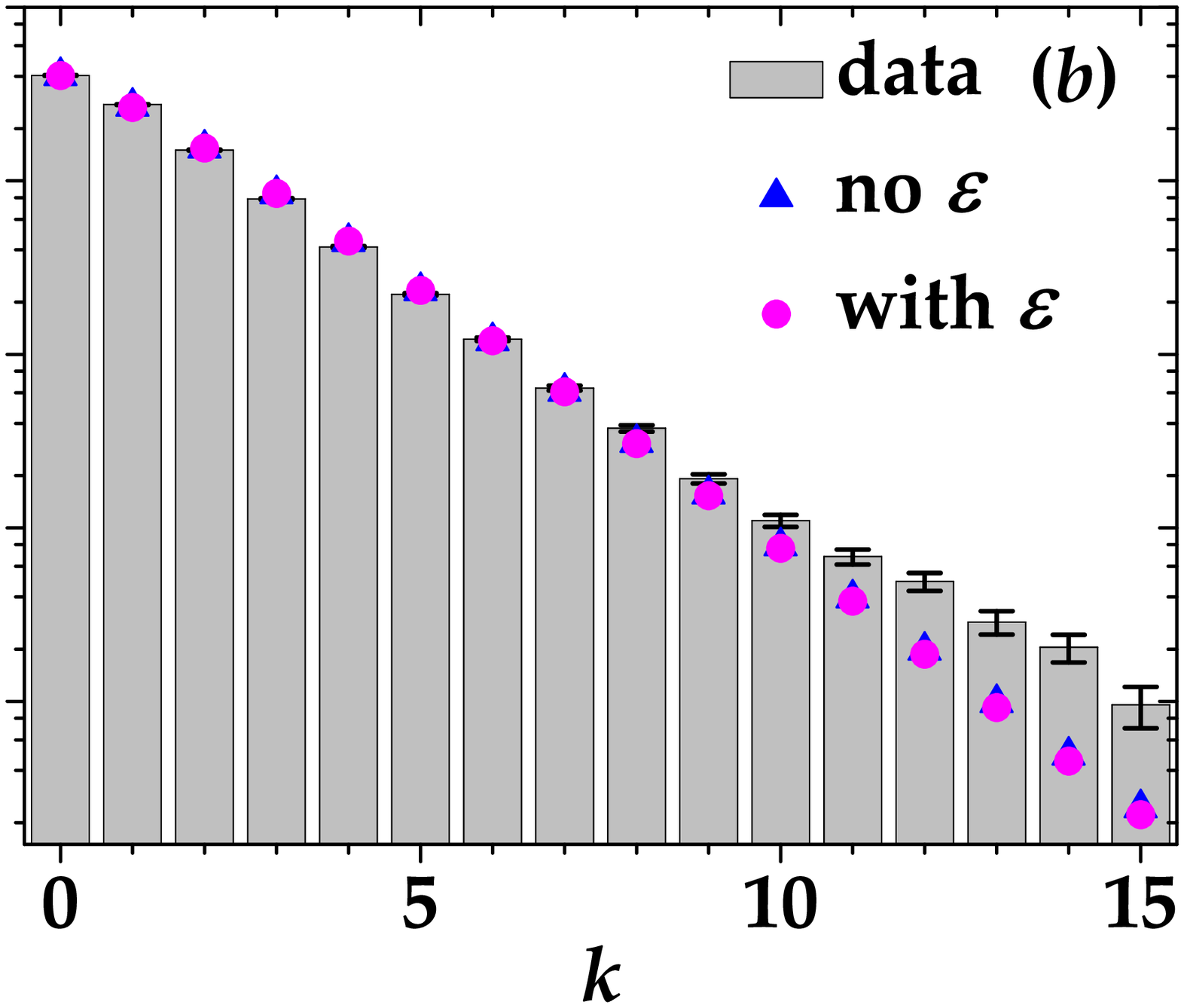}
\vspace{-0.5cm}
\end{center}
\caption{(Color online) Reconstructed P($k$) distributions in the case of multi-mode thermal light for 100-ns (a) and 50-ns (b) gate width. Gray columns + black error bars: experimental data; magenta dots: theoretical fitting curves in case the detector is affected by cross talk; blue triangles: theoretical curves in the absence of cross talk. In the presence of cross-talk effect, the values of  the number of modes are $\mu$ = 1.5148 $\pm$ 0.5387 and $\mu$ = 1.4790 $\pm$ 0.5201, respectively. The corresponding $\chi^2$ per degree of freedom is 14.24 in (a) and 14.79 in (b). On the contrary, in the absence of cross-talk effect we got $\mu$ = 1.3972 $\pm$ 0.0113 in panel (a) and $\mu$ = 1.3733 $\pm$ 0.0113 in panel (b). The corresponding $\chi^2$ per degree of freedom is 10.85 in (a) and 11.91 in (b). The contribution of dark counts is not considered in the theoretical fitting curves.}
\label{stboxtherm}
\end{figure}

The knowledge of all the moments of photon-number distribution can in principle provide much more information about the state under investigation. Even if the photon-number distribution represents only a portion, namely the diagonal, of the density matrix of an optical state, its knowledge is sufficient in many applications, such as  metrology \cite{zambra05}, material probing \cite{hofman}, biomedical optics \cite{gratton}, and optical communication protocols \cite{curty09}.\\
In this Section, we deal with the reconstruction of Poissonian statistics and of multi-mode thermal statistics by considering two choices (100 ns and 50 ns) of integration gate width.\\
For what concerns coherent light, 
%%%%%%%%%%%%%%%%%%%%%%%%%%%%%%%%
\begin{figure}[htbp]
\begin{center}
\hspace{-0.25cm}
\includegraphics[width=0.5\columnwidth]{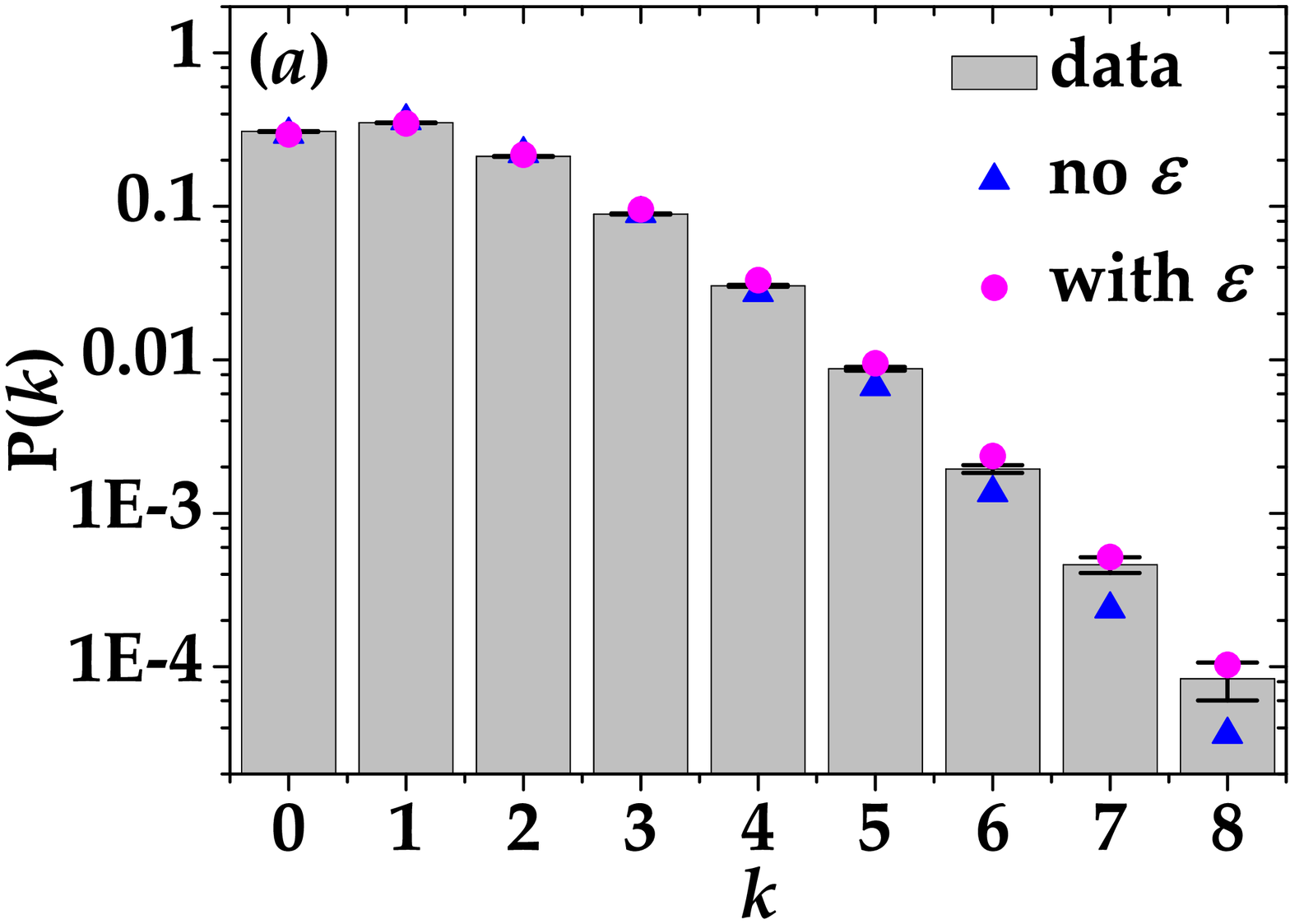}
\hspace{-1.6cm}
\includegraphics[width=0.5\columnwidth]{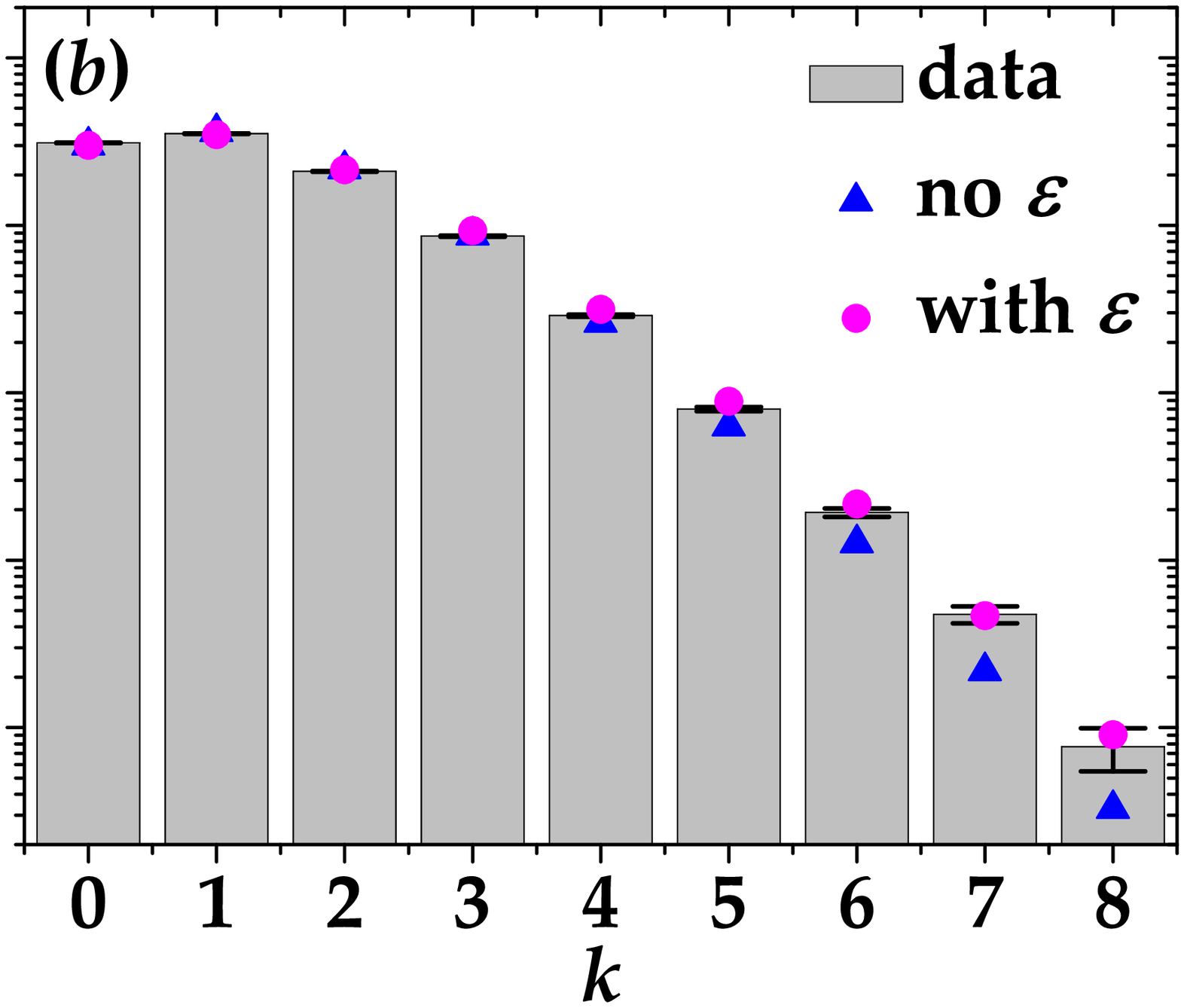} \\
\end{center}
\vspace{-0.5cm}
\caption{(Color online) Reconstructed P($k$) distributions in the case of coherent light for two choices of gate widths: 100 ns in panel (a) and 50 ns in panel (b). Gray columns + black error bars: experimental data; magenta dots: theoretical fitting curves in case the detector is affected by cross talk; blue triangles: theoretical curves in the absence of cross talk. The values of the $\chi^2$ per degree of freedom are: 34.47 and 28.02 in the absence of cross talk and 25.19 and 23.07 in the presence of it, respectively.}
\label{grstatcoher}
\end{figure}
%%%%%%%%%%%%%%%%%%%%%%%%%%%%%%%%
in Fig.~\ref{grstatcoher} we show the statistics of the actual signal amplitude $k$ on a logarithmic scale: 100-ns gate in panel (a) and 50-ns gate in panel (b). The mean value of the measured light was roughly the same in both cases, namely $\langle k \rangle = 1.3$. The data, shown as gray columns + black error bars, are plotted together with two possible fitting curves. The blue triangles represent the theoretical expectation in the case of coherent light,
\begin{equation} \label{Pcoher}
{\rm P_{coh}}(m) = \frac{\langle m \rangle^m}{m!} \exp{(-\langle m \rangle)},
\end{equation}
whereas the magenta dots take into account the occurrence of cross-talk effect at the detection stage \cite{ramilli}:
\begin{eqnarray}\label{PcoherXC}
{\rm P_{coh}}(k) &=& e^{- \langle m \rangle}(1 - \epsilon^{-k})\epsilon^k \frac{\sin(k \pi)}{k \pi}\\
& \times & _pF_q \left(1, -k; \frac{1}{2} - \frac{k}{2}, 1- \frac{k}{2};- \frac{1-\epsilon^2 \langle m \rangle}{4 \epsilon} \right), \nonumber 
\end{eqnarray}
where $_pF_q$ is the generalized hypergeometric function.
In Eqs.~(\ref{Pcoher}) and (\ref{PcoherXC}), the term $\langle m \rangle$ also includes the contribution of dark counts that cannot be discriminated from that of detected photons due to the light statistics.
In the theoretical expectation according to Eq.~(\ref{PcoherXC}), we used the values of $\epsilon$ shown in Tab.~\ref{tab1}. 
We note that, in each panel, the two theoretical curves are quite different from each other, especially for large numbers of $k$, and the complete theory that includes the cross-talk effect better fits the data. \\
%On the contrary, for smaller values of the gate (25 ns and 10 ns), the fitting curves are closely superimposed to each other and also to the experimental data. 
%The agreement between data and theory is also assessed by the values of fidelity, which are higher than 99.9$\%$ \cite{fidel}. 
%From these results we can argue that: (i) the shorter the gate the less important the cross-talk effects for the reconstruction of photon-number statistics; (ii) the shorter the gate the better the reconstruction.\\
As a second example of statistics reconstruction, in Fig.~\ref{stboxtherm} we plot P$(k)$ in the case of multi-mode thermal light with $\langle k \rangle = 1.3$ for the same choices of gate width shown above.
%%%%%%%%%%%%%%%%%%%%%%%%%%%%%%%%%%
%%%%%%%%%%%%%%%%%%%%%%%%%%%%%%%%%%
In more detail, the experimental data were obtained by integrating the signal over 100 ns in panel (a) and 50 ns in panel (b). 
In both cases the data are shown together with the theoretical fitting curves in the absence of cross talk (blue triangles)
\begin{equation}\label{Pth}
{\rm P_{mth}}(m) = \frac{(m+\mu-1)!}{m!(\mu -1)!(\langle m \rangle/\mu+1)^{\mu}(\mu/\langle m \rangle+1)^m}, 
\end{equation}
and in the presence of it (magenta dots)
\begin{eqnarray} \label{PthXC}
{\rm P_{mth}}(k) &=& e^{k} (1 - \epsilon)^{-k} \left( 1 + \frac{\langle m \rangle}{\mu + \epsilon \mu} \right) ^{- \mu} {{0}\choose{k}}\\
& \times & _pF_q \left(1, -k, \mu; \frac{1}{2} - \frac{k}{2}, 1- \frac{k}{2}; \frac{(-1+\epsilon)^2 \langle m \rangle}{4 \epsilon (\mu + \epsilon \mu + \langle m \rangle)} \right).\nonumber
\end{eqnarray}
Note that in Eq.~(\ref{PthXC}) we do not consider the contribution of dark counts. Indeed, the values of $\langle x \rangle_{\rm dc}$ shown in Tab.~\ref{tab2} are negligible with respect to the mean value of $k$. Thus, the term $\langle m \rangle$ represents the true mean number of detected photons.
In this case the models in the presence and in the absence of cross talk do not differ too much from each other (as also proved by the $\chi^2$ per degree of freedom reported in the caption of Fig.~\ref{stboxtherm}) and are well superimposed to the data up to a quite large number of photons (up to 10). The discrepancy for larger numbers is emphasized by the logarithmic scale, even if the absolute values are quite similar and almost negligible (of the order of $10^{-4}$).

\subsection{Correlations}
\begin{figure}[htbp]
\begin{center}
\includegraphics[width=0.8\columnwidth]{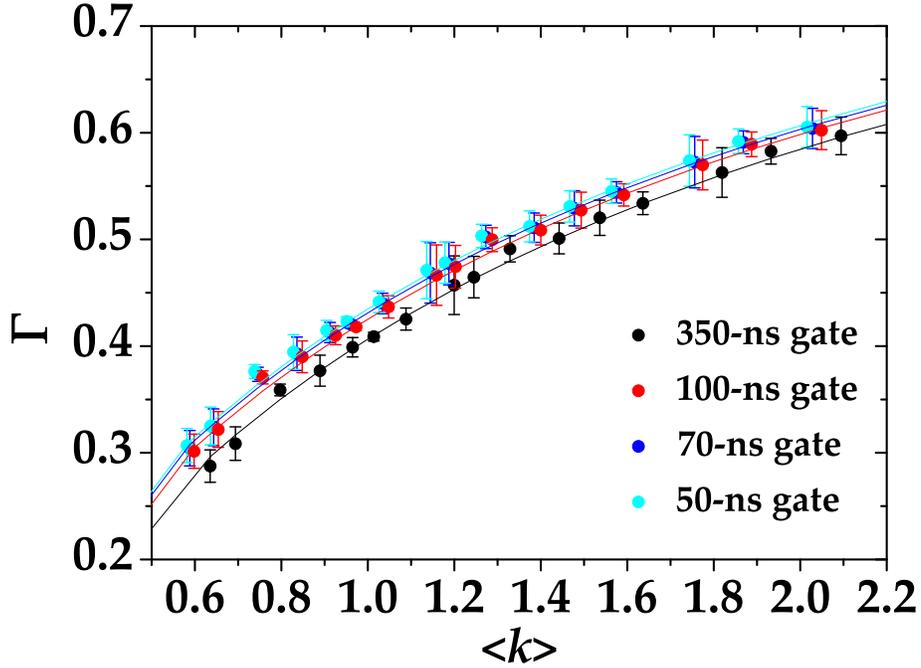}
\end{center}
\caption{(Color online) Experimental correlation coefficient $\Gamma$ (dots + error bars) as a function of $\langle k \rangle$ for pseudo-thermal light together with theoretical fitting curves (lines). Different colors correspond to different gate widths, namely 350 ns (black), 100 ns (red), 70 ns (blue), and 50 ns (cyan).
The fitting parameter is $\mu$, while $\epsilon$ is fixed from the plots in Fig.~\ref{FScoher} and $\langle m \rangle_{\rm dc}$ from the values reported in Tab.~\ref{tab2}. The fitting parameter obtained in the four cases is shown in Tab.~\ref{tab4}.}
\label{clcorr}
\end{figure}

The study of cross-correlations between the two parties of bipartite systems represents a powerful method to highlight the fluctuations of photon numbers, that goes beyond the direct reconstruction of the photon-number distributions. For instance, correlations have been used to emphasize the difference between a thermal statistics and a super-thermal one \cite{OL15,SciRep17}.
Moreover, they are of fundamental importance for many applications, such as for imaging \cite{ferri05,santa17,meda17,genovese} and generation of conditional states \cite{praIPS,lamperti}, both for classical and quantum states.\\ 
The calculation of photon-number correlations requires shot-by-shot determinations of the number of photons, a task that becomes harder and harder as the number of photons increases. In order to test the capability of SiPMs in properly revealing the presence of cross-correlations for different gate widths, we consider pseudo-thermal light divided at a balanced beam splitter.\\
Ideally, the shot-by-shot correlation coefficient for detected photons in the case of a multi-mode thermal state reads as:
\begin{equation}\label{corrmth}
\Gamma_{\rm mth} = \frac{\sqrt{\langle m_1 \rangle/\mu_1 \cdot \langle m_2 \rangle/\mu_2}}{\sqrt{(1 + \langle m_1 \rangle/\mu_1) \cdot (1 + \langle m_2 \rangle/\mu_2)}}. 
\end{equation}
Taking into account the presence of cross-talk effect and dark counts and using the relation between $\langle m \rangle$ and $\langle k \rangle = (\langle m \rangle + \langle m \rangle_{\rm dc}) (1+\epsilon)$, Eq.~(\ref{corrmth}) must be modified as follows \cite{OLcorr}:
\begin{widetext}
\normalsize
\begin{equation}\label{corrmthXCDCR}
\Gamma = \frac{\sqrt{{\left[ \langle m \rangle _1 /\left(1+\epsilon_1\right) -\langle m \rangle _{\rm dc,1} \right] \left[ \langle m \rangle _2 /\left(1+\epsilon_2\right) -\langle m \rangle_{\rm dc,2} \right]}{(\mu_1 \mu_2)^{-1}}}}{\sqrt{\left[1+\mu_1^{-1}\left( \langle m \rangle _1 / \left( 1+\epsilon_1 \right) -\langle m \rangle_{\rm dc,1} \right)\right] \left[ 1+\mu_2^{-1}\left(\langle m \rangle _2 / \left(1+\epsilon_2\right) -\langle m \rangle_{\rm dc,2} \right)\right]}}. 
\end{equation}
\end{widetext}
In Fig.~\ref{clcorr} we show the reconstructed shot-by-shot correlation coefficient for four different choices of gate widths, namely 350 ns (black dots), 100 ns (red dots), 70 ns (blue dots), and 50 ns (cyan dots). The experimental data are shown together with the fitting curves calculated according to Eq.~(\ref{corrmthXCDCR}), in which the values of cross-talk are those obtained from the evaluation of the Fano factor (see Tab.~\ref{tab1}). As to dark counts, we used the value reported in Tab.~\ref{tab2} for each gate width. The only left fitting parameter is then the number of modes $\mu$, which was assumed to be the same for the two outputs of the beam splitter. 
The values of $\mu$ are shown in Tab.~\ref{tab4} together with the values of the $\chi^2$ per degree of freedom.
%%%%%%%%%%%%%%%%%%%%%%%%%%%%%%%%%%%%
%%%%%%%%%%%%%%%%%%%%%%%%%%%%%%%%%%%%%
\begin{table}[!htb]
\begin{center}
\renewcommand{\arraystretch}{1.5}
\begin{tabular}{|c|P{1.5cm}|P{2.8cm}|P{1.5cm}|}
\hline 
gate width & $\mu$ & CI($\mu$) & $\chi^2_{\nu}$ \\
\hline 
350 ns & 1.3015 & (1.2876, 1.3154) & 0.28 \\ 
\hline 
100 ns & 1.2647 & (1.2543, 1.2750) & 0.34 \\ 
\hline 
70 ns & 1.2473 & (1.2370, 1.2576) & 0.30 \\ 
\hline 
50 ns & 1.2331 & (1.2230, 1.2432) & 0.38 \\ 
\hline 
\end{tabular} 
\caption{Values of the fitting parameter $\mu$ of the correlation coefficient as a function of the gate width in the case of multi-mode thermal light. The symbol CI indicates the 95$\%$ confidence interval. In the last column the $\chi^{(2)}$ per degree of freedom is shown.}\label{tab4}
\end{center}
\end{table}
%%%%%%%%%%%%%%%%%%%%%%%%%%%%%%%%%%%%%%%%%%%%%%%%%%%%%%%
We note that, given the same mean value $\langle k \rangle$, the highest values of $\Gamma$ correspond to the shortest gate width and thus to the smallest values of $\epsilon$ and $\langle m \rangle_{\rm dc}$. This behavior proves that the presence of cross-talk effect and dark counts decreases the amount of cross-correlation. Indeed, such effects independently occur in the two employed detectors.
However, it is worth noting that including the non-idealities is only necessary when the signal is integrated over long gate widths. On the other hand, the data corresponding to 50-ns gate could be well fitted by a curve in which both dark counts and cross-talk effect are set equal to zero. This result confirms that for the employed SiPMs the correction given by Eq.~(\ref{corrmthXCDCR}) to Eq.~(\ref{corrmth}) is substantially negligible, at least for small gate widths.

\subsection{A further improvement}

\begin{figure}[htbp]
\begin{center}
\includegraphics[width=0.8\columnwidth]{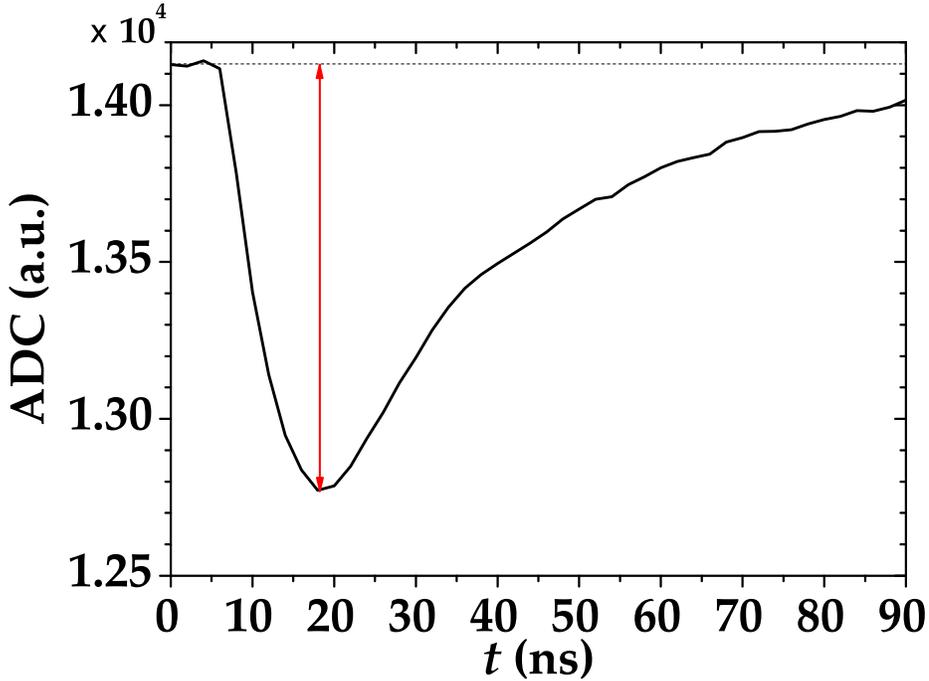}
\vspace{-1cm}
\end{center}
\caption{(Color online) A single-shot detector signal, acquired with the peak-and-hold circuit, is indicated as black curve. The red arrow indicates that in this case the height of the peak was extracted shot-by-shot.}
\label{oscilloscope1}
\end{figure}

All the results achieved so far indicate that the shorter the gate the most negligible the non-idealities are. However, the integration over short gates can rise an issue: The setting of the gates is quite delicate, since a precise control of temporal delays is required. In order to optimize the acquisition of the light signal and make the spurious stochastic effects negligible, we designed a low-noise shaping amplifier. When this device is connected to a SiPM output, the typical output signal appears as in Fig.~\ref{oscilloscope1}. 
%%%%%%%%%%%%%%%%%%%%%%%%%%%%%
\begin{figure}[htbp]
\begin{center}
\includegraphics[width=0.8\columnwidth]{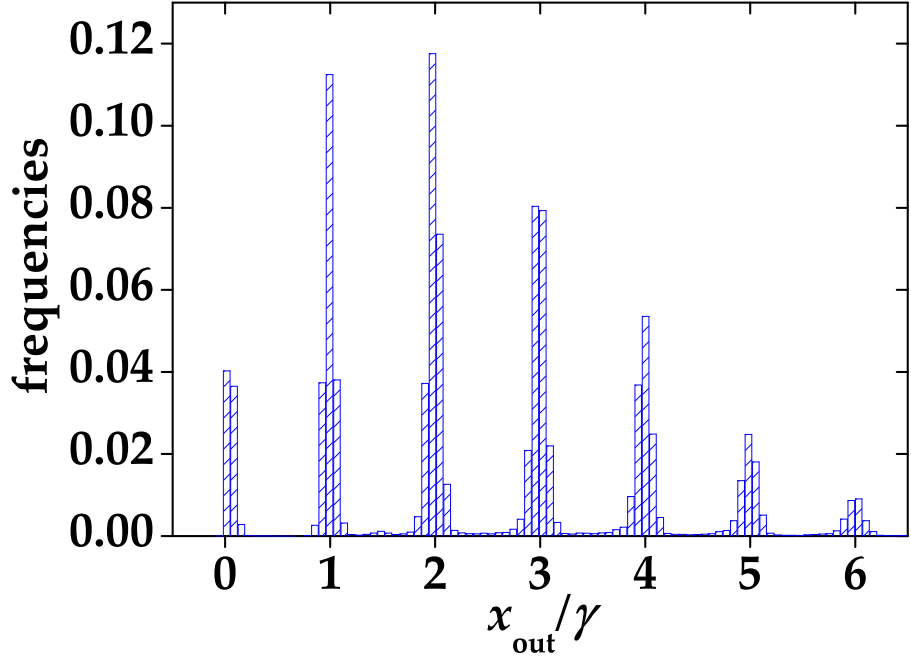}
\end{center}
\caption{(Color online) Normalized pulse-height spectrum corresponding to a coherent state with $\langle m \rangle = 2.56$, obtained by applying the peak-and-hold acquisition system to the signal output.}
\label{PHS}
\end{figure}
Due to this shaping, it is possible to implement a full digital peak capture procedure with the DT5730 digitizer, that gets rid of the non simultaneous spurious effects. 
The difference between this digitizer and the one described in the previous Sections is the sampling rate, which is twice as large (500 MS/s).
In Fig.~\ref{PHS} we show a typical pulse-height spectrum after the SiPM detected a coherent light beam, where a good peak separation is appreciable. It is worth noting that, at variance with the pulse-height spectra shown in Fig.~\ref{PHSd}, the peak corresponding to 0 photons in Fig.~\ref{PHS} is asymmetric and its distance from the 1-photon peak is different from the typical peak-to-peak distance.
The reason for this bias is due to the specific analysis procedure, which consists in evaluating shot-by-shot the height of the peak. Since also in the absence of light the maximum of the signal is calculated, the 0-photon peak is not centered in 0, but rather translated in the positive direction.  
\begin{figure}[htbp]
\begin{center}
\hspace{-0.25cm}
\includegraphics[width=0.5\columnwidth]{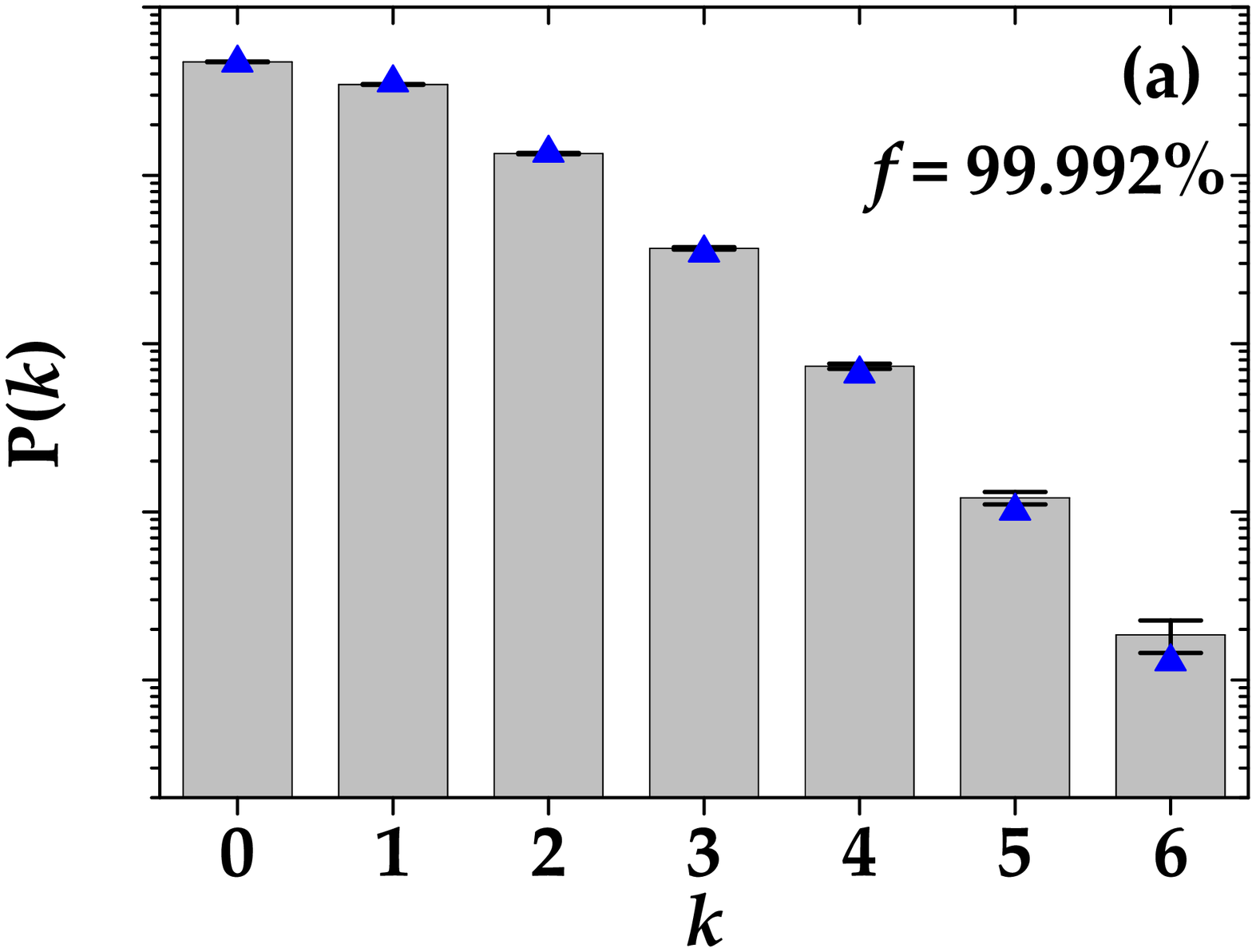}
\hspace{-1.6cm}
\includegraphics[width=0.5\columnwidth]{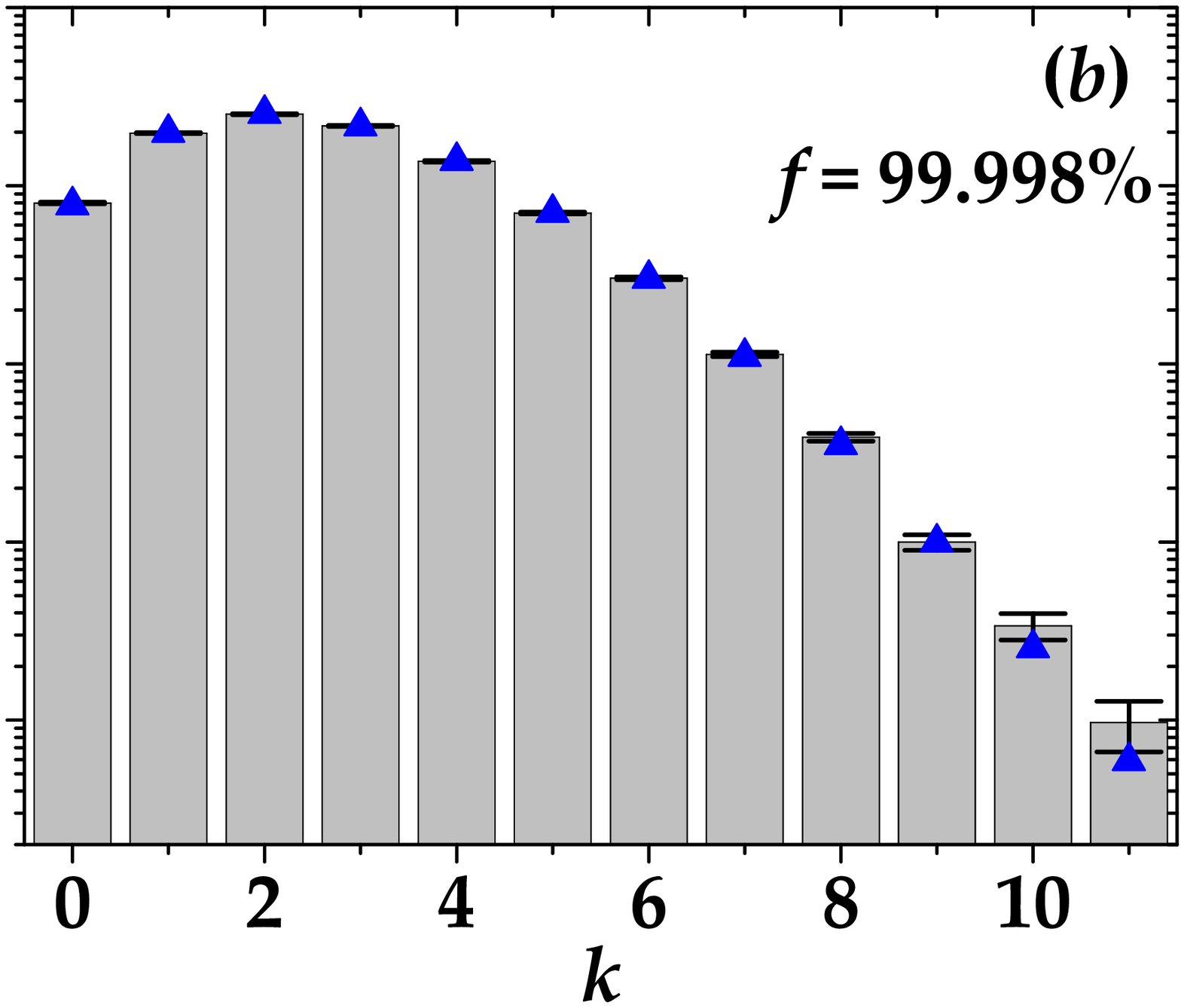}
\vspace{-0.5cm}
\end{center}
\caption{(Color online) Reconstructed P($k$) distributions in the case of coherent light acquired with the peak-and-hold circuit to the output signal. (a): Coherent light with $\langle m \rangle = 0.76$; (b): Coherent light with $\langle m \rangle = 2.56$. Gray columns + black error bars: Experimental data; blue triangles: Theoretical curves in the absence of cross talk. The corresponding $\chi^2$ per degree of freedom is 8.31 in (a) and 1.33 in (b).}
\label{histo}
\end{figure}
In this case, the S/N corresponding to Fig.~\ref{PHS} is calculated as the ratio betweeen the mean value of the 1-photon peak and its width. In particular, its value is S/N = 13 $\pm$ 1.2, which is lower than the maximum value shown in Fig.~\ref{SN}. However, the peak-and-hold circuit allows for a better reconstruction of the statistical properties since the stochastic effects are essentially negligible. On the contrary, in the case of integration over a specific gate width we have seen that such effects play an important role modifying the statistical properties of light. 
Indeed, the reconstructed photon-number distributions shown in Fig.~\ref{histo} do not require to consider non-idealities to match the theoretical expectations. 
This configuration encourages the exploitation of the new generation of SiPMs in the Quantum Optics context.

\section{Conclusions}
We presented a thorough analysis of the capability of the new generation of SiPMs produced by Hamamatsu to properly reconstruct the statistical properties of classical states of light. We demonstrated that we are able to describe the detector outputs by including in our theoretical model cross-talk effect and dark counts. We pointed out how, under proper experimental conditions, these drawbacks can be neglected. In fact we proved that the influence of these effects on the measurement of light states depends on the integration gate, thus making the peak-and-hold acquisition strategy preferable. Indeed, by using a peak-and-hold acquisition system matched with a proper front end electronics, dark counts can be neglected, as, so far, the optical cross talk.  
Even if further improvements are still needed, such as a better beam coupling into fibers, the employed detection chain paves the way towards the exploitation of SiPMs in more complex scheme. For instance, this kind of detectors could be used in the homodyne-like detection scheme based on hybrid photodetectors (HPDs) that we recently applied to coherent-state discrimination \cite{bina} and to quantum-state reconstruction \cite{qtomo}. Since SiPMs are more compact than HPDs, they would allow the portability of the apparatus, thus increasing the number of possible applications fields.

\end{document}